\documentstyle[aps,prd,psfig]{revtex}
\begin{document}

\title{ Can There be Quark Matter Core in a Strongly Magnetized \\
Neutron Star?}
\author{Tanusri Ghosh$^{a)}$\thanks{E-mail:tanusri@klyuniv.ernet.in}
and Somenath Chakrabarty$^{a,b)}$\thanks{E-mail:somenath@klyuniv.ernet.in}\\
 a) Department of Physics, University of Kalyani, Kalyani 741 235,
India\thanks{Permanent address}\\
b) Inter-University Centre for Astronomy and Astrophysics, Post Bag 4\\
Ganeshkhind, Pune 411 007, India}

\date{\today}
\maketitle
\begin{flushleft}
\end{flushleft}

\begin{abstract}

The effect of strong quantizing magnetic field on the nucleation of quark matter
droplets and on the chemical evolution of nascent quark phase at the core of
a neutron star are investigated. The surface energy of quark phase diverges
logarithmically. As a consequence there can not be a first order transition
to quark phase. However, a metal-insulator type of second order transition is 
possible unless the field strength exceeds $10^{20}$G. The study of chemical
evolution of newborn quark phase shows that in $\beta$-equilibrium the system 
becomes energetically unstable.

\end{abstract}
\bigskip\bigskip

The theoretical investigation of the effect of ultra-strong magnetic field on 
stellar matter
in nuclear astrophysics has got a new dimension after the discovery of 
a few magnetars.The observed soft gamma repeaters discovered by BATSE 
\cite{bat} and KONUS \cite{kon}  (see also \cite{hu})
experiments and X-ray sources observed by ASCA, RXTE and BappoSAX \cite{as}
show the presence of 
strong surface magnetic field  up to $10^{15}$G. The discovery of these objects 
pose a great challenge to the existing models of magnetic field evolution, 
since they require a very rapid field decay in isolated neutron stars 
\cite{col}. To 
investigate the global properties of these strange objects, it also requires a 
detail investigation of stability of dense stellar matter in presence of 
ultra-strong magnetic field and know the exact equation of state of such 
strongly magnetized matter. 

The dynamo mechanism, recently  proposed  by Thompson and Duncan \cite{th}
suggests that 
the dipole magnetic field of a young neutron star can reach up to $10^{15}$G. 
It is generally expected that the internal magnetic field is a few orders of 
magnitude stronger than surface field strength.
Since the strength of internal magnetic field of a neutron star 
strongly depends on the nature of dense stellar matter present at the 
core region, it may not necessarily be reflected in its surface magnetic 
field. However, there is an upper limit of internal magnetic field strength 
constrained by the scalar virial theorem, which gives 
$B_{\rm{max}}\sim 2\times 10^8 (M/M_\odot)(R/R_\odot)^{-2}$G \cite{lai,sh}. 
For a typical neutron  star of radius $R=10$km and mass $M= M_\odot$, 
this upper limit is $\sim 10^{18}$G. Beyond this limit, the 
ultra-magnetized neutron stars become unstable.

Now there is also a strong belief that a transition to quark phase 
occurs at the core of a neutron star if the density exceeds a few times 
normal nuclear density. The transition could be a first order, with the 
nucleation of stable quark matter droplets in metastable hadronic matter 
by fluctuation at the core region.  The transition  could also be a second 
order type. 
Since hadrons (neucleons and hyperons) do not carry color quantum number, 
we may call the hadronic matter  a color insulator. Whereas, in this 
regard,  the quark phase is a color conductor. Therefore, such a second 
order structural phase transition at high density is analogous to the 
metal insulator transition in condensed matter physics, which takes place 
under high pressure.

In this letter our aim is to show that a first order transition to 
quark phase is absolutely forbidden at the core of a young pulsar if 
the magnetic field strength exceeds $10^{15}$G for which the Landau 
levels of the quarks are populated.
On the other hand a second order transition is possible unless the 
magnetic field strength exceeds $10^{20}$G. Now the transition to quark 
phase in a young strongly magnetized neutron star is a strong interaction 
process (time scale $\sim 10^{-23}$sec.). The constituents (up ($u$), 
down ($d$) and strange ($s$) quarks and electrons) of the 
quark phase are however not in $\beta$-equilibrium immediately after their 
formation. The chemical equilibrium time scale (weak interaction time scale) 
is much longer than the droplet nucleation time scale. In this context 
we shall further like to show that if in the quark phase  the magnetic 
field strength exceeds the typical value $4.4\times 10^{13}$G, for which 
the Landau levels of electrons are populated \cite{sc1,sc2,sc3,sc4},  
in the   chemical or 
$\beta$ equilibrium condition quark phase becomes energetically unstable 
with respect to the corresponding hadronic phase.

Let us consider the nucleation of stable quark matter droplets  due to 
fluctuation in a metastable hadronic matter. The rate of nucleation is given by
\cite{lan}
\begin{equation}
I=I_0 \exp \left (- \frac{\sigma^3}{C} \right )
\end{equation}
where $I_0$ and $C$ are finite constants, $\sigma$ is the surface tension 
of the quark phase formed in metastable hadronic matter. The detail 
analytical structure of these constants are not important in our study. 
Then considering the MIT bag model of color confinement, we have the 
expression for surface tension of the quark phase in presence of a 
strong quantizing magnetic field of strength $B$
\begin{equation}
\sigma_i=\frac {3TB}{8\pi} q_i \sum_{\nu=0}^\infty \int_0^\infty 
\frac{dk_z}{(k_z^2+k_\perp^2)^{1/2}} \ln \left [ 1+ \exp 
\left [ -\frac{\epsilon_{\nu (i)}-\mu_i}{T} \right ] \right ] G
\end{equation}
where $i$ indicates the species (up, down or strange quarks), $q_i$ is 
the flavor charge, $k_{\perp (i)}=(2\nu q_i B)^{1/2}$ is the transverse 
momentum, $\nu$  is the Landau quantum number, 
$\epsilon_{\nu (i)} =(k_z^2 +k_{\perp (i)}^2 +m_i^2 )^{1/2}$ is the 
modified form of single particle energy, $m_i$ and $\mu_i$ are the mass 
and chemical potential respectively and $G=1-2\tan^{-1}(k/m_i)/\pi$.

Whereas in the non-magnetic or non-quantizing magnetic field case \cite{ol,be}
(see also \cite{ja}),
\begin{equation}
\sigma_i=\frac {3T}{32\pi^2}  \int \frac{d^3k}{k} \ln \left [ 1+ \exp 
\left [ -\frac{\epsilon_i-\mu_i}{T} \right ] \right ] G
\end{equation}

Now it is just a matter of simple integration by parts to show from eqn.(2) 
that the surface tension of the quark phase diverges logarithmically for 
$\nu=0$ (zeroth Landau level). The surface tension goes as $\sim -\ln(\nu)$ 
as $\nu \rightarrow 0$ in presence of a quantizing magnetic field, whereas 
the surface tension as given in eqn.(3) for the non-magnetic case is a finite 
quantity. Therefore the rate of nucleation of quark matter droplets as given 
in eqn.(1) becomes identically zero. Which concludes that if the magnetic 
field strength at the core of a neutron star is strong enough to populate 
Landau levels of the quarks the nucleation of quark droplets become 
impossible, which means that under such strange condition  a first order 
transition to quark phase is absolute forbidden.

However, a second order phase transition to quark matter can not be ruled 
out. This is a continuous transition. The surface tension of the new phase 
has no role in the process. The chemical potential, density and pressure 
change continuously at the phase boundary. 
Using these conditions at the phase boundary, we have obtained numerically 
the critical density  for the phase transition as a function of magnetic 
field strength. In fig.1 we have plotted this variation. This figure shows 
that the critical density (solid curve)
diverges at $B \approx 10^{20}$G, which is of course 
too high to achieve at the core of a newborn neutron star. However, for 
relatively lower values of magnetic field strength, the critical densities 
are finite and well within the limit of central density of a stable neutron 
star. Therefore, a metal-insulator kind second order transition is possible 
even in presence of a strong magnetic field of astrophysical interest.
The bag parameter also becomes unphysical beyond $B\approx 10^{18}$G (shown
by dashed curve).

As we have already mentioned that  such a transition takes place in the 
strong interaction time scale. Therefore, the produced quark phase will not 
be in $\beta$-equilibrium. It takes comparatively longer time ($\sim$ weak 
interaction time scale) to achieve chemical equilibrium in the system.

To study the chemical evolution of the system, we consider the following 
weak processes in quark matter phase: 
$d\rightarrow u+e^- +\bar{\nu_e}~(1),~ u+e^- \rightarrow d+\bar{\nu_e} ~(2),~ 
s\rightarrow u+e^- +\bar{\nu_e} ~(3),~ u+e^- \rightarrow s+\bar{\nu_e} ~ (4),  ~ 
u+d \leftrightarrow u+s ~(5).$
We have further assumed that the neutrinos are non-degenerate (they leave the 
system immediately after their formation). We have noticed that the trapping of 
neutrinos (if the phase transition occurs at the core of a proto-neutron star) 
in the quark phase do not change our conclusions. 
The approach to chemical equilibrium  of the system is therefore governed 
by the following sets of kinetic equations
\begin{eqnarray}
\frac{dY_u}{dt}&=& \frac{1}{n_B}[\Gamma_1- \Gamma_2+\Gamma_3-\Gamma_4]\\
\frac{dY_d}{dt}&=& \frac{1}{n_B}[-\Gamma_1+ \Gamma_2-\Gamma_5^{(d)}+
\Gamma_5^{((r)}]
\end{eqnarray}
where $n_B$ is the baryon number density, $Y_i=n_i/n_B$ is the fractional 
abundance of the species $i$ and $\Gamma_j$'s are the rates of the 
processes $j=1,2,3,4,5$. The indices $d$ and $r$ are respectively for the 
direct and reverse processes for  $j=5$.
The baryon number conservation and charge neutrality conditions  
give $Y_s=3-Y_u-Y_d$ and $Y_e=Y_u-1$ respectively.
For a neutron star of mass $\approx 1.4M_\odot$, the baryon number density at 
the centre is $3-4$ times normal nuclear density, temperature $\sim 10^9$K and 
proton fraction is about $4\%$. Then the initial conditions are $Y_u(t=0)=1.04$,
$Y_d(t=0)=1.96$. As a consequence of baryon number conservation and charge 
neutrality, we have $Y_s(t=0)=0$ and $Y_e(t=0)=0.04$.
Solving numerically the kinetic equations along with the conditions of baryon 
number conservation and charge neutrality and using the initial conditions as 
given above, we have  obtained the time variation of fractional abundances for 
various species
for a given baryon number density. In fig.2 we have plotted the time dependences
of fractional abundances for
various species in the  zero magnetic field case. In this situation the 
fractional abundances for up, down and strange quarks saturate to their 
$\beta$-equilibrium values, whereas the electron fraction becomes extremely 
small. In fig.3 we have plotted the same variations when only electrons are 
affected by quantizing magnetic field ($B=10^{14}$G). In this case the 
strangeness fraction first increases but ultimately both the down and strange 
quark abundances go to zero and the system effectively behaves like a up-quark 
core in the $\beta$-equilibrium condition. For 
the sake of charge neutrality the electron fraction also becomes very 
high in this particular physical situation in the 
equilibrium condition.
In fig.4 we have shown the time evolution of the species when the magnetic 
field strength is strong enough ($B= 5\times 10^{15}$G) to affect electrons 
and also up and down quarks. In this case the strange quarks are never produced.
In the $\beta$-equilibrium condition, therefore, the main constituents are up 
and down quarks with a small fraction of electrons.

Now the phase transition to quark matter may occur at the core of a neutron 
star from dense hyperonic matter (with non-zero initial strangeness fraction)
instead of pure nucleonic matter. 
In this case the initial strange quark abundance is finite. We have noticed 
that the qualitative nature of the curves do not change even if the initial 
hyperon fraction is $30\%$.
In this particular scenario, in the field free case the fractional abundances
of all the three quarks saturate to $\beta$-equilibrium values (as shown in 
fig.2). If only electrons are affected, the variation of strangeness fraction 
and down quark abundance are exactly identical with fig.3. Although, the 
initial strangeness fraction is non-zero, it ultimately vanishes in the
$\beta$-equilibrium condition. Same is the case when the magnetic field is 
strong enough to affect  electrons as well as up and down quarks. The 
equilibrium values for the  fractional abundances changes only $\approx 5\%$ 
with respect to the corresponding values when the initial strangeness 
fraction is zero.

To study the stability of the quark matter system under various physical 
situations discussed above, we have calculated energy per baryon in the 
$\beta$-equilibrium condition for different baryon number densities. Under 
zero pressure condition, the energy per baryon is given by
\begin{equation}
\epsilon= \frac{1}{n_B} \sum_{u,d,s,e}\mu_i n_i
\end{equation}
We have obtained energy per baryon as a function of baryon number density for 
all the cases discussed in the text, including the hypothetical 
situation-an almost 
flavor symmetric quark matter in an external magnetic field (according to our 
calculation such a situation can not be achieved in the real world). In fig.5 
we have shown the variation of energy per baryon with baryon number density 
for various physical scenarios. For the sake of comparison we have plotted 
the stability curve of neutron matter with Bethe-Johnson equation of state 
(dashed curve) \cite{sh}. From the nature of the curves we can conclude that the 
hypothetical quark matter scenario is the most stable configuration. The field 
free quark matter with and without degenerate neutrinos are also energetically 
stable up to 2-2.5 times normal nuclear density. This is  consistent
with the speculation of Witten \cite{wi}. Whereas, in all other physical 
situations, in presence of a strong magnetic field the quark matter is 
energetically unstable.

Therefore the final conclusions are  (i) if the magnetic field strength 
exceeds $10^{15}$G, a first order transition to quark phase
is absolutely forbidden at the 
core of a neutron star. (ii) A metal-insulator kind of second order 
transition is however possible unless the field strength is $>10^{20}$G, 
which is of 
course too high to have  at the core of a neutron star. (iii) Even if the 
transition is of second order in nature, because of two kinds of completely
different time scales, the chemical evolution of the system 
in presence of a strong magnetic field ($\geq 4.4 \times 10^{13}$G) leads to 
energetically unstable configuration. Therefore quark matter core is absolutely
impossible in an young pulsar with strong magnetic field. However, in very 
old neutron stars with much weaker magnetic field strength such a phase 
transition is possible by matter accretion which may increase the density of 
the core.

If the situation is such that a pulsar is very young and strongly magnetized 
and at the same time compact enough, then probably pion or / and kaon 
condensation will play the major role to make the matter energetically stable
\cite{sah}.
%

\begin{figure} 
\psfig{figure=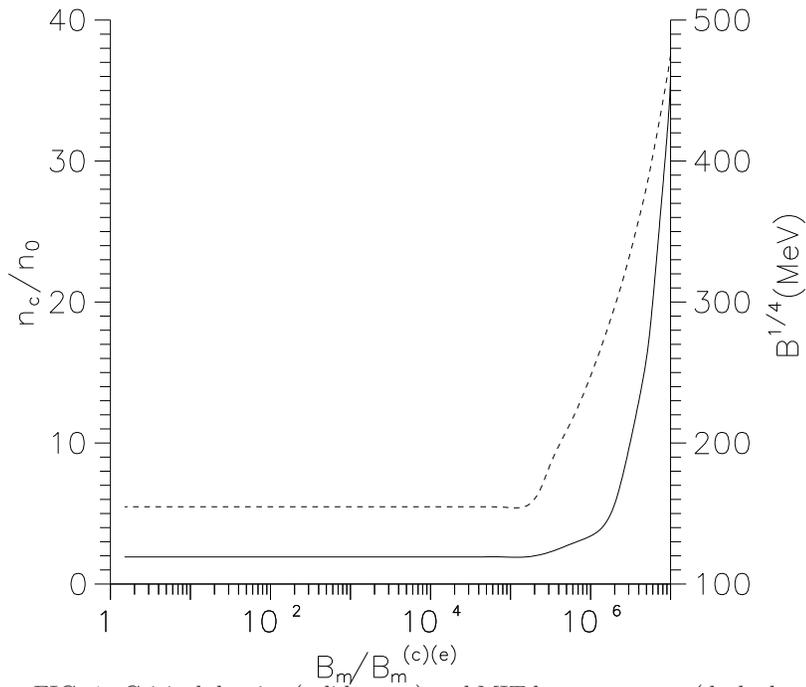,height=0.5\linewidth}
\caption[]{Critical density (solid curve)  and MIT bag parameter (dashed curve)
for metal-insulator type second order transition as a function of magnetic 
field strength. The magnetic 
field strength is expressed in terms of the critical field for electron}
\end{figure}
\begin{figure} 
\psfig{figure=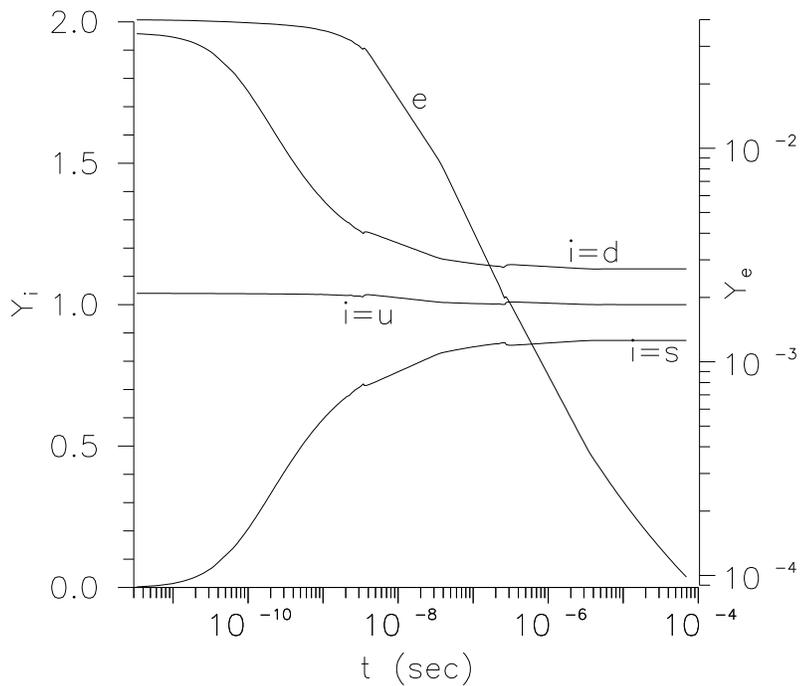,height=0.5\linewidth}
\caption[]{Fractional abundances for various species in the  zero magnetic field
case.}
\end{figure}
\begin{figure} 
\psfig{figure=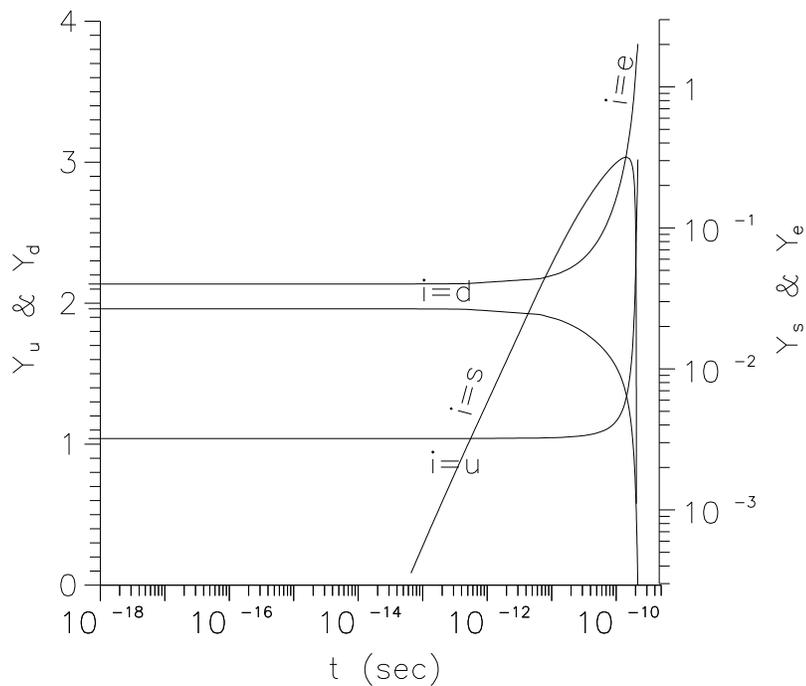,height=0.5\linewidth}
\caption[]{ Same as fig.2 when only electrons are affected by quantizing 
magnetic field ($B=10^{14}$G). }
\end{figure}
\begin{figure} 
\psfig{figure=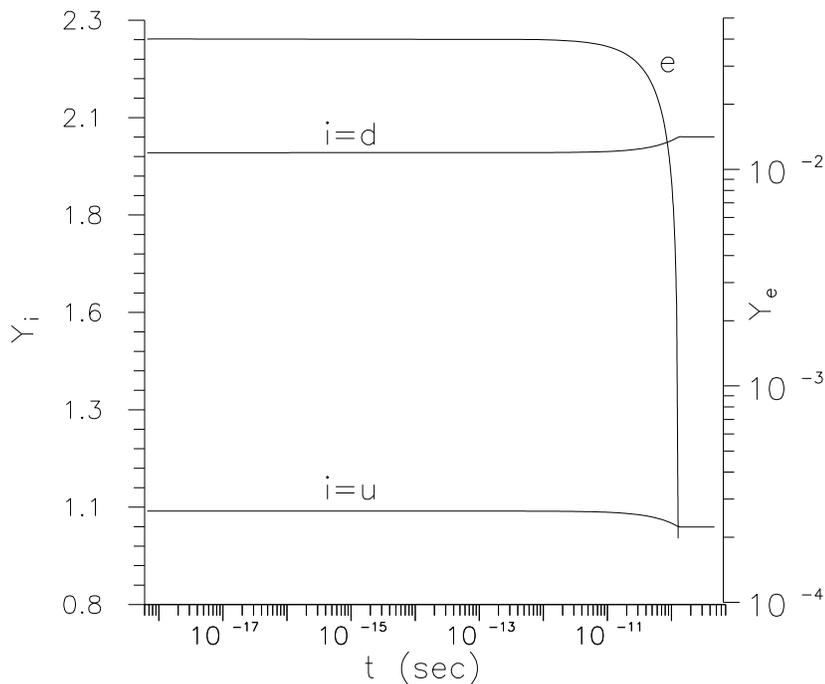,height=0.5\linewidth}
\caption[]{ Same as figs.2 and 3 but the magnetic field is strong enough to
affect electrons as well as up and down quarks ($B=5 \times 10^{16}$G). }
\end{figure}
\begin{figure} 
\psfig{figure=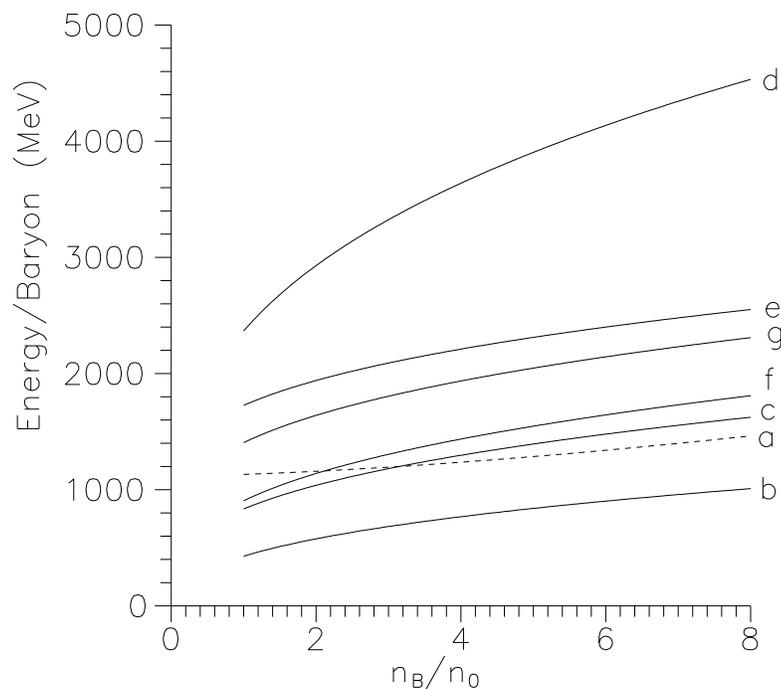,height=0.5\linewidth}
\caption[]{Variation of energy per baryon for different physical
conditions (see text). Curves (a) is for neutron matter, (b) is for 
hypothetical quark matter placed in an external magnetic field, (c) is for 
quark matter
in the field free case,
(d) is for quark matter in presence of a strong magnetic field $B_m=10^{14}$G,
(e) is same as (d) for $B_m=5\times 10^{16}$G, (f) is for neutrino
trapped quark matter in the field free case and (g) is same as (f) for $B_m=5
\times 10^{16}$G.}
\end{figure}
\end{document}